# Solution growth of Ce-Pd-In single crystals: characterization of the heavy-fermion superconductor $Ce_2PdIn_8$

Klára Uhlířová, Jan Prokleška, Vladimír Sechovský, Stanislav Daniš

Department of Condensed Matter Physics, Faculty of Mathematics and Physics, Charles University in Prague, Ke Karlovu 5, 12116 Prague 2, Czech Republic

**Abstract**
Solution growth of single crystals of the recently reported new compound $Ce_2PdIn_8$ was investigated. When growing from a stoichiometry in a range 2:1:20 - 2:1:35, single crystals of $CeIn_3$ covered by a thin (~50 μm) single-crystalline layer of $Ce_2PdIn_8$ were mostly obtained. Using palladium richer compositions the thickness of the $Ce_2PdIn_8$ layers were increased, which allowed mechanical extraction of single-phase slabs of the desired compound suitable for a thorough study of magnetism and superconductivity. In some solution growth products also $CePd_3In_6$ ($LaNi_3In_6$ –type of structure) and traces of phases with the stoichiometry $CePd_2In_7$, $Ce_{1.5}Pd_{1.5}In_7$ (determined only by EDX) have been identified.
Magnetic measurements of the $Ce_2PdIn_8$ single crystals reveal paramagnetic behaviour of the $Ce^{3+}$ ions with significant magnetocrystalline anisotropy. Above 70 K the magnetic susceptibility follows the Curie-Weiss law with considerably different values of the paramagnetic Curie temperature, for the magnetic field applied along the *a*- (-90 K) and *c*-(-50 K) axis. Below the reported critical temperature for superconductivity $T_c$ (0.69 K) the electrical resistivity drops to zero. Comparative measurements of the electrical resistivity, heat capacity and AC susceptibility of several crystals reveal that the superconducting transition is strongly sample-dependent.

## 1. Introduction

In some Ce-based compounds, the interaction of the Ce ions with the conduction electrons often leads to a large enhancement of the effective electron mass. These so-called heavy-fermion (HF) compounds exhibit attractive electronic properties, such as strongly enhanced paramagnetism, non-Fermi liquid behaviour, interplay between magnetism and superconductivity (SC), etc. During the last decade, the attention of many researchers has been focused on the series of HF materials of the general chemical formula $Ce_mT_nIn_{3m+2n}$ where $m, n$ = 1 or 2 and $T$ = Co, Rh, Ir. These compounds crystallize in the $Ho_mCo_nGa_{3m+2n}$-type tetragonal structures with the space group P4/*mmm*. These structures are built by $n$ layers of distorted cuboctahedra [$CeIn_3$] and one monolayer of rectangular parallelepipeds [$TIn_2$] stacked sequentially in the [001] direction [1] making the structure to be quasi-2D. The main building block of these compounds, the cubic $CeIn_3$, is a heavy-fermion antiferromagnet (AF) with a Néel temperature $T_N$ = 10.2 K [2] and becomes superconducting (SC) under pressure [3]. The whole $Ce_mT_nIn_{3m+2n}$ series is characterized by interplay between magnetism and unconventional superconductivity, which makes these compounds suitable for thorough studies of varieties of the two cooperative phenomena in strongly correlated electron systems. The quasi-2D $R_mT_nIn_{3m+2n}$ crystal structure provides investigation of the effect of varying dimensionality on magnetism and unconventional superconductivity because the structures become less 2D-like with increasing $n$. The respective compounds are often called 115 ($m$, $n$ = 1) and 218 ($m$ = 2, $n$ =1) compounds.

CeCoIn$_5$, CeIrIn$_5$ and Ce$_2$CoIn$_8$ [4-7] are ambient-pressure superconductors, CeRhIn$_5$ and Ce$_2$RhIn$_8$ are antiferromagnetic (AF), tuned to the SC phase by applying pressure or by doping [8-10]. In a certain pressure range, coexistence of AF and SC has been observed. While the superconducting transition temperature $T_c$ is increasing with increasing applied pressure the Néel temperature $T_N$ is decreasing and antiferromagnetism vanishes when $T_N = T_c$. In the range of pressures where $T_N < T_c$, the magnetism became hidden and can be recovered by magnetic field [11]. On the other hand the superconductivity can influence development of field-induced magnetic order as observed for CeCoIn$_5$ [12].

Most recently the group of Ce$_n$$T$In$_{3n+2}$ was extended to Ce$_2$PdIn$_8$. Ce$_2$PdIn$_8$ was first reported by Shepa *et al*. [13], while an independent study has shown a non-Fermi liquid behaviour [14] and HF SC where the SC emerges out of the long-range AF state at $T_c$ = 0.68 K and coexist with the AF at ambient-pressure [15]. Based on our results and because the reported Néel temperature equals to the Néel temperature of CeIn$_3$, we have ascribed the AF origin to the intergrowth of CeIn$_3$ phase into Ce$_2$PdIn$_8$ [16].

Last year, Kurenbaeva *et al*. [17] reported a new structure type, CePt$_2$In$_7$, with a tetragonal structure (space group I4/*mmm*) formed by two layers of [PtIn$_2$] and one layer of [CeIn$_3$] connecting this compound with the Ce$_n$$T$In$_{3n+2}$ family. According to recent results presented by E. D. Bauer [18], this compound should be another Ce-based superconductor.

Both the 115 and 218 compounds were mostly prepared in single crystalline form using the solution growth technique [1, 4, 19]. Unfortunately, in many cases the description of sample preparation is limited by citing a source where only general information about the method is presented, or even by unpublished work. Although this might be a strategy of the authors, we would like to show that the sample quality and its history might be a crucial parameter for the subsequent physical behaviour. In this paper the SC of single-crystalline Ce$_2$PdIn$_8$ will be presented together with detailed description of sample preparation.

## 2. Experimental
*2.1. Crystal growth of Ce$_2$PdIn$_8$*
The Ce-Pd-In phase diagram as reported by Shepa *et al*. [13] is rather rich and according to our results probably still not completed. Unfortunately there are many compounds existing in rather broad Ce-Pd and Pd-In solubility ranges, which complicates the sample synthesis namely when high quality of the samples is required.
The solution growth technique [20, 21] was used for the synthesis of single crystal samples. The high purity alumina (99.8 %) crucibles or, for the best-found growth conditions, ultrahigh purity alumina (99.99 %) crucibles by *Almath crucibles ltd*. have been used to reach the maximum sample purity.

In the first experiments, the single crystals of Ce$_2$PdIn$_8$ were grown from In flux using the starting stoichiometry ratio of Ce, Pd and In 2:1:20-30. The starting elements of high purity (In – 99.999 %, Pd – 99.95 %, Ce – 99.9 %) were put into alumina crucibles and sealed in quartz glass under high vacuum. After that the samples were put into a programmable furnace and a thermal process was started; the system was heated up to 950 °C hold there for 120 minutes and than slowly (2-4 °C/min) cooled down to 350-400 °C. The crucibles were than replaced into new quartz glass ampoule with a quartz-wool stopper and evacuated again to heat it up to 350 °C at which the remaining indium was centrifuged.

While crystals of the other $Ce_2TIn_8$ ($T$ = Co, Rh, Ir) compounds can be prepared rather easily, the case of growing $Ce_2PdIn_8$ is rather complicated but also even more interesting. Numerous discrepancies were observed during the characterization process; while the microprobe analysis from the sample surface confirmed the $Ce_2PdIn_8$ composition, the X-ray powder diffraction (XRPD) resulted mostly in $CeIn_3$. It was found that after growing out of the above mentioned stoichiometry, multiphase products were formed. Mainly cuboid-shaped single crystals of $CeIn_3$ covered by a very thin layer (50-100 μm) of $Ce_2PdIn_8$ were obtained. This lead us to the idea that $CeIn_3$ grew until part of cerium was consumed, than, in a narrow concentration region, $Ce_2PdIn_8$ was grown and finally, from the reaming palladium in the melt, $Pd_3In_7$ was formed.

Based on these results, a Pd-richer composition was used to suppress the initial growth of $CeIn_3$. Many experiments in a broad concentration range of Pd and In have been done following the estimation of products ratio in previous experiments. The best composition has been found to be $CePd_{2-3}In_{35}$. From the higher Pd content, $CePd_3In_6$ ($LaNi_3In_6$ [22]) was formed covering the surface of $Ce_2PdIn_8$ (Supplementary Fig. 5). The details of characterization of $CePd_3In_6$ will be published elsewhere. The pure $Ce_2PdIn_8$ has not been reached yet but the $Ce_2PdIn_8$ layers in the $CeIn_3$-$Ce_2PdIn_8$ sandwiches were thick enough (~ 200 μm) to be mechanically separated. The border between the two phases is very well defined and sharp as demonstrated in Fig. 1 and Supplementary Figs. 1-3. The element mapping Fig.1 (left) shows a cut polished surface of the so called $CeIn_3$ - $Ce_2PdIn_8$ sandwich. The line scan in Fig. 1 (right) presents the relative intensity of the Ce and Pd spectra along the marked arrow. A minority of a third phase with the composition determined from the EDX analysis as $Ce_{1.5}Pd_{1.5}In_7$ was detected. In some cases it appeared that a "multilayered" system with rather thick (50-200 nm) well defined layers of $CeIn_3$ and $Ce_2PdIn_8$ was formed as presented in supplementary Fig. 5. In some batches, traces of $CePd_2In_7$ were detected by EDX analysis. One should be also careful of inclusions of $Pd_3In_7$ in $Ce_2PdIn_8$ single crystals.

Since $CeIn_3$ oxidizes much faster than $Ce_2PdIn_8$ the boundaries between these phases are observable by naked eye or optical microscope (Supplementary Fig. 1). The Ce-oxide on the $CeIn_3$ phase makes also the back scattered electron (BSE) contrast more clear. The BSE contrast between $CeIn_3$ and $Ce_2PdIn_8$ at just polished (no oxide) samples is hardly visible because palladium and indium have similar atomic masses and difference in volume densities of both compounds (7.820 $g.cm^{-3}$ and 8.062 $g.cm^{-3}$ for $CeIn_3$ and $Ce_2PdIn_8$, respectively) is only 3%. The separation process is described in Supplementary information.

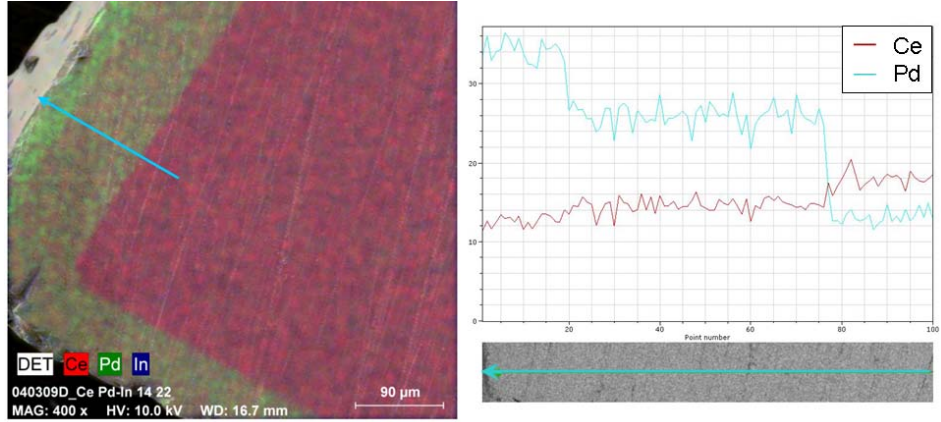

Fig. 1. Element mapping (left) of the cut and polished CeIn$_3$-Ce$_2$PdIn$_8$ sandwich. CeIn$_3$ (central, red) is covered by layer of Ce$_2$PdIn$_8$, a small region of new phase with nominal composition Ce$_{1.5}$Pd$_{1.5}$In$_7$. The line scan (right) along the blue arrow show a sharpness of the boundary between phases

*2.2 Characterization and measurement techniques*

The samples were checked by X-ray powder diffraction (XRPD) on a Bruker D8 Advance diffractometer, and by a scanning electron microscope (SEM) Tescan Mira I LMH equipped by an energy dispersive X-ray detector (EDX) Bruker AXS.

Magnetic properties, resistivity and heat capacity were measured on MPMS and PPMS (using $^3$He option) devices (Quantum Design). The AC magnetic susceptibility ($\chi_{AC}$) at low temperatures (0.35<T<2.5K) was measured using custom made extension to the PPMS apparatus allowing to measure $\chi_{AC}$ using the ACMS option with the $^3$He insert (including the ACMS preamplifier). The self-made coil set-up with two counter-wounded secondary coils (Cu wire) and primary coil (NbTi wire) was used for this measurement. Details about the set-up, necessary modification and test measurements are described elsewhere [23].

The low temperature part of resistivity was measured in a magnetic field of 0.03 T, because in zero magnetic field a SC transition of indium was detected (a kink around 3.2 K).

All measured samples were plate like single-crystals with a thickness ~100 μm and a mass of ~1.5 mg; the *c*-axis was perpendicular to the surface. The measurements were done on several crystals of the composition Ce$_2$PdIn$_8$ (confirmed by EDX). For this work, three representative crystals, marked as the sample "A", "B" and "C", respectively, were chosen.

**3. Results**

By the XRPD the Ho$_2$CoGa$_8$-type of structure was confirmed, the lattice parameters have been determined as: $a$ = 0.4695 nm and $c$ = 1.220 nm. The parameter *a* is almost equal to the lattice parameter of the cubic CeIn$_3$ ($a$ = 0.4689 nm) which supports the stability of this system. On the other hand, the lattice parameter $a$ = 0.4691 nm of Ce$_2$IrIn$_8$ matches the one of CeIn$_3$ even better and no evidence of forming such sandwiches has been observed.

Magnetic measurements on $Ce_2PdIn_8$ single crystals show paramagnetic behaviour with significant magnetocrystalline anisotropy. The temperature dependence of the reciprocal DC susceptibility ($1/\chi$), presented in Fig. 2, is linear above ~70 K, i.e. the $\chi$ ($= M/H$) vs. $T$ dependence of the susceptibility follows the Curie-Weiss law with $\mu_{eff} = 2.6$ $\mu_B/Ce^{3+}$, $\Theta_p = -90$ K, and $\mu_{eff} = 2.6$ $\mu_B/Ce^{3+}$, $\Theta_p = -50$ K for the magnetic field $B\|a$ and $B\|c$, respectively. At lower temperatures the crystal field effects become important and the magnetic susceptibility departs from the Curie-Weiss law; in the $c$-direction, the magnetic susceptibility departs from the C-W behaviour at about 70 K and become constant below 23 K, while for the $a$-direction it departs form the C-W behaviour at 22 K and became almost constant below 15 K, see Fig. 2 (inset). Below 8 K, there is an increase of the susceptibility in both directions. Such behaviour is usual for many Ce-compounds and is frequently ascribed to a paramagnetic contribution of other rare earth elements presented in the Ce metal in a very small amount (< ~100 ppm). Of course, it can also signify a phase transition in the temperature region below 2 K, but no anomaly has been observed in resistivity and heat capacity data.

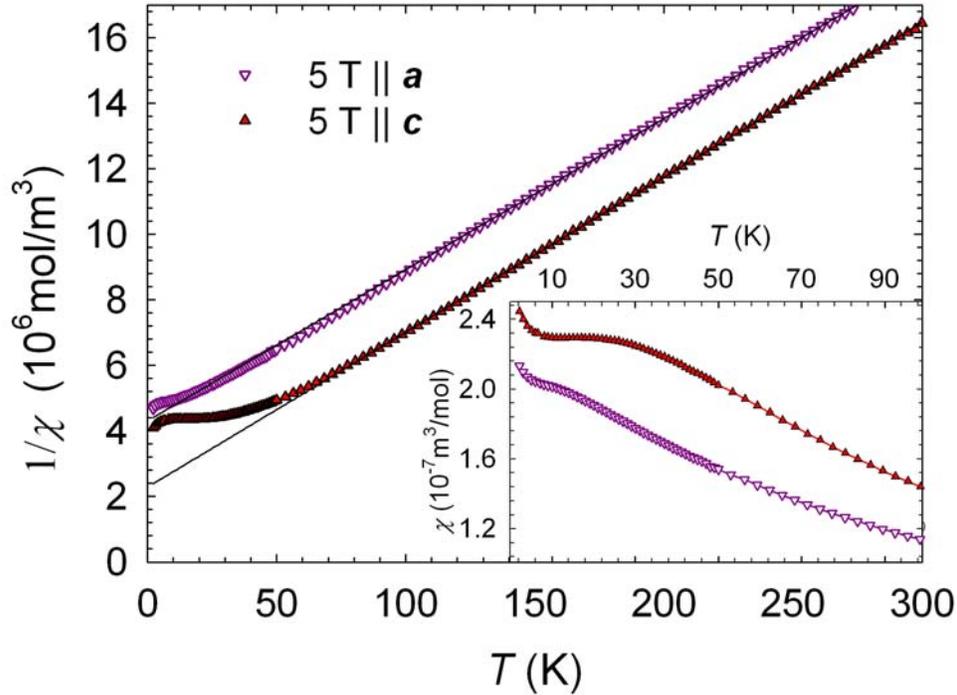

Fig. 2. Temperature dependence of reciprocal magnetic susceptibility of $Ce_2PdIn_8$ measured along both principal crystallographic directions, black lines are the C-W law fits. The inset shows a low temperature part of the magnetic susceptibility.

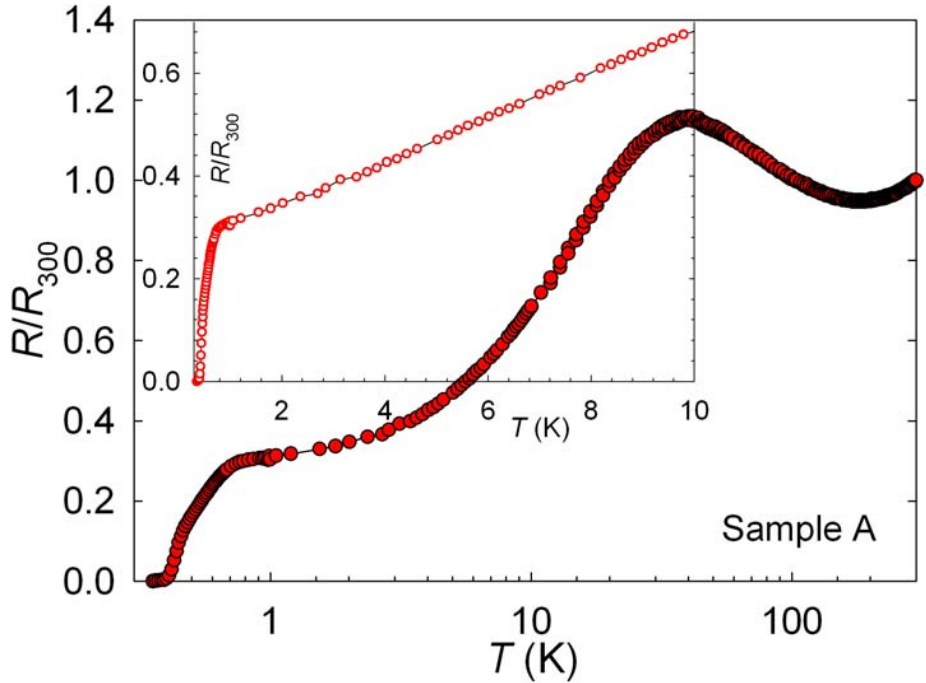

Fig. 3. Temperature dependence of electrical resistivity of $Ce_2PdIn_8$. The low temperature part (inset) was measured in magnetic field of 0.03 T to suppress the SC transition of indium.

In Fig. 3 the temperature dependence of the resistivity measured with an AC current applied in the basal plane is presented. A maximum is evident at the temperature $T_{max} \sim 40$ K, this is the typical behaviour of other members of this family of compounds (more pronounced in 218's [9, 10, 24]). The maximum is caused by the magnetic part of the resistivity and corresponds to onset of low-temperature coherent behaviour of a Kondo lattice, and the characteristic temperature is generally marked as $T^*$ [25]. A rough estimation of $T^*$ can be also obtained from the departure of the susceptibility data from the Curie-Weiss behaviour [26] yielding the value $T^{*\chi} = 40$ K (polycrystalline average). Although both values are in agreement, we would be careful to draw conclusion, since they were found to be sample dependent. The presented values were determined from sample "A" whereas e.g. sample "B" shows to the values of $T^* \sim 30$ K.

Below $\sim 10$ K a linear temperature dependence of the resistivity was observed (Fig. 3 inset) until the resistivity drops to zero value manifesting the SC transition (Fig. 5). Linear dependence of resistivity confirms the non-Fermi liquid behaviour reported in [14], which may indicate that the compound appears on the verge of magnetism and the superconductivity is magnetically driven. Below critical temperature $T_c < 0.69$ K a SC transition was observed. The critical temperature was determined by resistivity, heat capacity and AC magnetic susceptibility measurements see Figs. 4-6. As presented in Fig. 6 and Fig. 7, the heat capacity and AC susceptibly measurements detected more SC transitions. The highest value of $T_C = 0.69$ K (determined from the temperature derivative of the AC susceptibility) was observed in sample "C"; however heat capacity and AC magnetic susceptibility data evidenced another transition at 0.46 K in this sample. The latter transition was detected in all measured samples with various intensity. A weak pronounced heat capacity anomaly at ~ 0.46 K was responsible for the broad resistivity decrease to zero in sample "A"; however the main transition in sample "A" seems to be below 0.36 K as hinted in Fig. 6 (inset). The last case is sample "B", for which three anomalies were observed, the resistivity showed obviously a drop to zero value around the highest critical temperature.

The discrepancies in the low temperature behaviour of $Ce_2PdIn_8$ were also observed by Kaczorowski *et al.* [14,15], where for the polycrystalline sample they presented normal paramagnetism down to the lowest reached temperature $T \sim 0.35$ K, while the single crystal was superconducting below $T_c = 0.68$ K. Passing over the other presented (magnetic) transitions above $T_c$, which are of controversial origin, they detected one clear SC phase transition both by resistivity and heat capacity measurements, the highest $T_c$ observed for $Ce_2PdIn_8$.

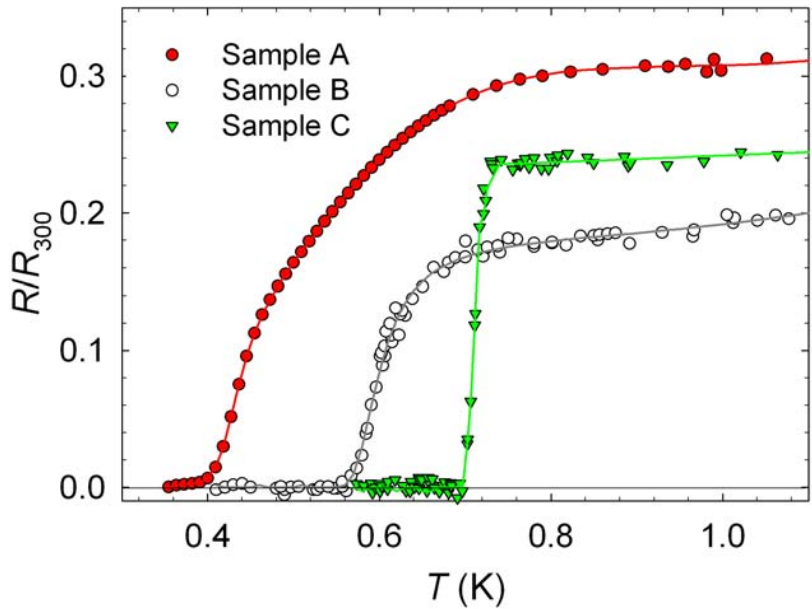

Fig. 4. Low temperature region of electrical resistivity of three different samples of $Ce_2PdIn_8$ (A, B and C) showing on sensitivity on sample quality (measured at magnetic field of 0.03 T).

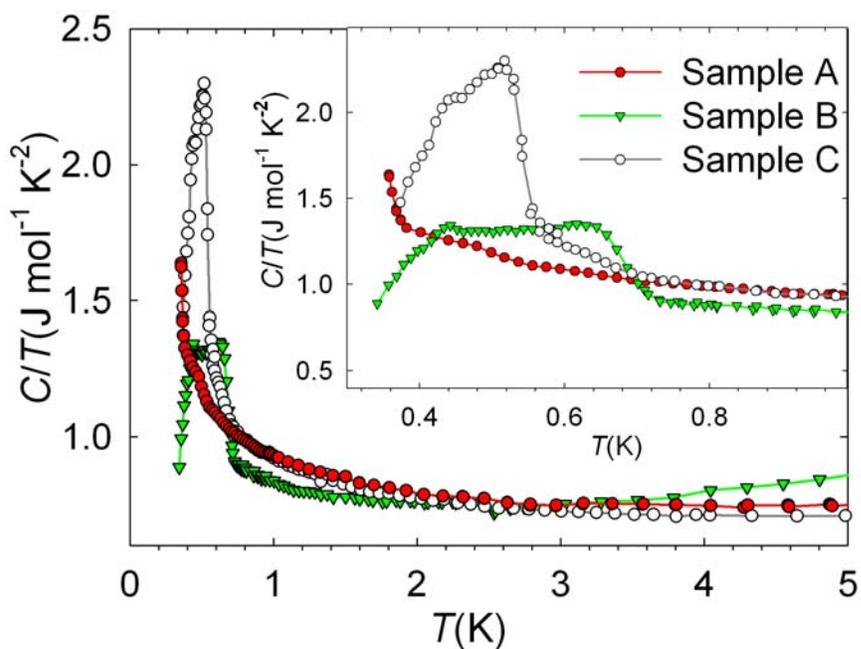

Fig. 5. Low temperature region of heat capacity of the samples A, B and C.

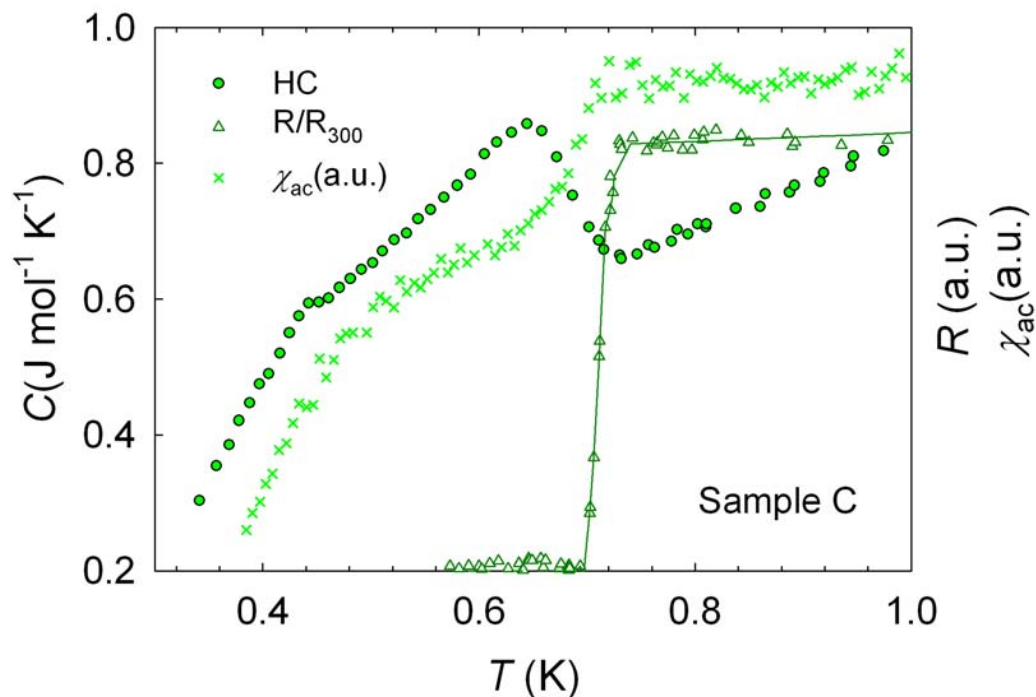

Fig. 6. Heat capacity, resistivity and real part of AC magnetic susceptibility for sample B.

## 4. Conclusions

Although we have not found optimal conditions for the growth of the $Ce_2PdIn_8$ single crystal, we were able to separate single-phased samples for basic physical measurements.

The study of $Ce_2PdIn_8$ showed that the compound is paramagnetic down to the SC transition. The SC transition was found to be sample dependent; the highest $T_c$ value has been determined to be 0.69 K.

Our samples were carefully (faces and edges of the plate-like crystals) checked by EDX analysis. Nevertheless, also this method has its limits and therefore we cannot 100 % exclude a presence of small amount of impurity phase or small variation of composition in our samples below the sensitivity of the EDX detector. With respect to the magnitude of the effects we exclude the influence of (completely different) impurity phase and ascribe the differences in the low-temperature behaviour to structural defects such as stacking faults typical for layered systems or small variation of composition or both. This is motivated by results on $Ce_2RhIn_8$ [19] where high resolution neutron diffraction revealed $Ce_2RhIn_8$ to be "a complex mixture of two layered mosaic, polytypic phases affected by non-periodic partially correlated planar defects". Also high-resolution electron microscopy may provide important data for resolving these questions. Although the EDX analysis is not able to detect a possible small variation in the composition, it has proven to be a crucial tool for exploring this attractive system, especially when identifying and separating $CeIn_3$ and other spurious phases.

Considering that samples with the highest $T_c \sim 0.69$ K have the sharpest SC transition (resistivity measurement) and the same value of $T_c$ has been confirmed by AC susceptibility and heat capacity measurements (and also by [15]) one might guess that it is a property of the "best" $Ce_2PdIn_8$ crystal, while structural defects or deviation of stoichiometry shift the $T_c$ to lower values. Further studies of the sample behaviour depending on sample fabrication such as effects of cooling rate or annealing are necessarily to be done.

From our point of view, $CeIn_3$ and $Ce_2PdIn_8$ form a beautiful sandwich-like system; however, for physical property studies isolated well defined single crystals would be preferred. On the other hand, the evident tendency for forming such layered systems might be interesting for epitaxial grow studies as reported e.g. for $CeCoIn_5$ [27].

We have also synthesized new compounds, $CePd_3In_6$ and traces of $Ce_{1.5}Pd_{1.5}In_7$ and $CePd_2In_7$. Detailed characterization of these materials will be an object of our further studies; the results will be published elsewhere.

**Acknowledgments**

This work is a part of the research plan MSM 0021620834 that is financed by the Ministry of Education of the Czech Republic and has been also partly supported by the Czech Science Foundation grants no. 202/09/1027, 202/09/H041 and by the Charles University Grant Agency grant no. 109907. We would like to thank to our colleague Jiří Pospíšil for his help with building our solution-growth laboratory.